# Charge trapping and super-Poissonian noise centers in a cuprate high-temperature superconductor

K. M. Bastiaans[1]*, D. Cho[1]*, T. Benschop[1], I. Battisti[1], Y. Huang[2], M.S. Golden[2], Q. Dong[3], Y. Jin[3], J. Zaanen[1], M. P. Allan[1]¶

1 Leiden Institute of Physics, Leiden University, Niels Bohrweg 2, 2333 CA Leiden, The Netherlands

[2] Van der Waals–Zeeman Institute, Institute of Physics, University of Amsterdam, 1098 XH Amsterdam, The Netherland

[3] Centre de Nanosciences et de Nanotechnologies, CNRS, Univ. Paris-Sud, Univ. Paris-Saclay, C2N – Marcoussis, 91460 Marcoussis France

*These authors contributed equally to this work.  ¶ allan@physics.leidenuniv.nl

**The electronic properties of cuprate high temperature superconductors in their normal state are very two-dimensional: while transport in the ab plane is perfectly metallic, it is insulating along the c-axis, with ratios between the two exceeding $10^4$.[1–4] This anisotropy has been identified as one of the mysteries of the cuprates early on[5–7], and while widely different proposals exist for its microscopic origin[5,8–10], little is known empirically on the microscopic scale. Here, we elucidate the properties of the insulating layers with a newly developed scanning noise spectroscopy technique that can spatially map not only the current but also the current fluctuations in time. We discover atomic-scale noise centers that exhibit MHz current fluctuations 40 times the expectation from Poissonian noise, more than what has been observed in mesoscopic systems.[11] Such behaviour can only happen in highly polarizable insulators and represents strong evidence for trapping of charge in the charge reservoir layers. Our measurements suggest a picture of metallic layers separated by polarizable insulators within a three-dimensional superconducting state.**

The difference between metals and insulators is that in the former, additional charge equilibrates in femtoseconds while the latter can be statically charged. The coupling to the lattice is a necessary condition for the trapping of charge in insulators on macroscopic timescales since electrons by themselves are too quantum mechanical to localize. The trapped charge is stabilized by the formation of static, localized polarons involving a reconfiguration of the atomic lattice. Strongly polarizable insulators such as $SrTiO_3$ exemplify this process. Such trapping of charge on slow timescales has been conjectured as a cause of the anomalous, highly resistive c-axis transport in the cuprate high-temperature superconductors,[8] competing with alternative proposals such as incoherent transport[12,13]. We note that in these materials, bands structure theory predicts metallic transport along the c-axis due to the small but finite c-axis bandwidth of the order of 0.1eV[14]. Here, we present firm evidence that such charge trapping processes do occur in a superconducting cuprate material by measuring atomically resolved current fluctuations.

Quite generally, fluctuations in time of a signal - the noise - can be a powerful diagnostic tool as they contain information not present in the mean value. It has historically allowed to distinguish between signals carried by particles and waves or between black body radiation and coherent radiation of a laser. More recently, noise spectroscopy has established itself as a standard method of investigating mesoscopic systems. This is usually done by looking at the noise spectral power, $S(w) = <\delta I(t) \delta I(t')>$, where $\delta I$ is the deviation of the current operator from the mean, and the averaging $<>$ is both quantum mechanical and statistical. Examples where noise transport measurements led to novel insight include: the study of fractional charge[15,16], the doubling of charge in Andreev processes[17], and the vanishing of noise in break junctions at the quantum conductance[18]. We have succeeded in bringing this technique to the atomic scale in the tunnelling regime, discovering an unanticipated phenomenon when we applied it to a cuprate high temperature superconductor.

The flow of classical, uncorrelated charged particles between two leads is a purely Poissonian process. Its noise is independent of frequency (white) and proportional to the charge $q$ and the flow $I$ of the carriers, $S = q|I|$, as a direct consequence of the discreteness of the carriers.[19] We define the normalized noise $S_n = S/2e|I|$, similar to the Fano factor. Thus $S_n = 1$ represents Poissonian noise, $S_n < 1$ and $S_n > 1$ refer to sub-Poissonian and super-Poissonian noise, respectively (Fig. 1a-b). For an uncorrelated electronic liquid, one expects $S_n = 1$; for an exotic, spatially inhomogeneous electronic liquid, this is a priori unclear. Importantly, in order to find $S_n > 1$, processes are required in the system on a frequency scale that fits into the slow (MHz) frequency window of the noise measurements. Another property of special relevance to our data is that such noise events may be completely invisible in the mean value of the current.

Our aim is thus to measure the fluctuations in the cuprates on the atomic scale. Bringing noise measurement to the tunnelling regime comes with unique challenges which prevented any atomic resolution shot noise measurement thus far. The central obstacle lies in the high impedance of the tunnelling junction, which is typically 0.3 GOhm to 10 GOhm. Together with the capacitance between tip and sample and the cable capacitance, the junction acts as low pass filter only allowing transmission of signals in the kHz range where $1/f$ noise and mechanical resonances dominate. Possible solutions include bootstrapping of a amplifier[20] or building a impedance matching circuit.[21] Matching a GOhm junction leads to considerable losses in the circuit. This is simplified when using the STM in point contact mode or in the low MOhm range[21,22], however, this increases interactions between tip and sample, making it more difficult to extrapolate the sample properties. Our goal is to perform noise measurements when in tunnelling regime, with transparencies $t$~$10^{-6}$. To accomplish this task, we build a resonance circuit with all inductors from superconducting Nb, and include a custom-built[23], high mobility amplifier directly into the circuit, following the principle of devices built for noise spectroscopy measurements in mesoscopic systems.[24] Figure 1c shows the amplification circuit that allows us to map out noise in the MHz range with GOhm junction resistances. We thoroughly tested our setup on a gold sample (supplementary information).

We choose to first investigate with this new scanning noise microscopy instrument the cuprate high-temperature superconductors[25] with the hope to find signs of the slow, glassy fluctuations associated with charge- and/or current loop order that have been claimed to show up in the noise[26–28]. These are not present and instead we found a surprise that we will now explain.

As a sample material, we decided to use the slightly overdoped bi-layer cuprate $(Pb,Bi)_2Sr_2CaCu_2O_{8+\delta}$ with $T_c$ = 79K. The Pb substitution for Bi has the advantage of supressing the characteristic supermodulation seen in many Bi based cuprates, simplifying the interpretation and making higher voltage measurements possible. We cleave the samples in UHV at pressures below $10^{-10}$ mbar, and directly insert them into our STM head at 3.2 K. The tunnelling process starts with electrons originating from the CuO layer which then tunnel through the SrO and Bi layers[29]. Tunnelling through these charge reservoir layers does not have much effect on the STM signal, except for some spatial filtering.[29,30] Figure 2a shows a topographic image revealing atomic resolution. The square Bi-lattice is clearly resolved with some bright protrusions induced by Pb-substitutions for Bi atoms [31].

Upon recording the noise as a function of spatial location with atomic resolution at bias voltages of ±0.1 eV we find homogeneously Poissonian noise (with an accuracy below 2%), with no sign of the fluctuating orders. However, this changes dramatically when increasing the bias, as shown in Figure 2b-c. While most locations still exhibit Poissonian noise, a few atomic locations reveal striking enhancements of the noise. These super-Poissonian noise centers show noise values up to 40 times the expectation from Poissonian processes, more than anything that has been observed in mesoscopic systems.[11] We emphasise that the increase in noise is extremely localized, the width of the peaks in space being around 0.5 nm for most centers. The density of super-Poissonian noise centers is roughly 0.3% referenced to the number of Cu atoms, and they are evenly distributed throughout the sample.

A key insight comes from the energy dependence of the noise. Noise spectra are shown in Figure 2d-f. Below a certain threshold, the noise in the tunnelling current is purely Poissonian, $S_n$ = 1. But above around 1eV or below around -650meV the normalized noise rises rapidly. Surprisingly all noise centers appear to be highly asymmetric in energy: the noise enhancements at positive bias do not spatially correlate with the noise enhancement seen at negative energy.

While different locations show different strengths of noise enhancement, the onset energy seems to be independent of the noise center, roughly -0.8eV for the negative energy noise centers, and roughly +1eV for the positive ones. This indicates a common mechanism and turns out to be an important diagnostic tool, as discussed below.

To understand these observations, it is worth taking a step back and look at engineered and natural systems that exhibit non-Poissonian noise, $S_n \neq 1$. In electronic systems this is most often sub-Poissonian noise[11,18,20,32], usually due to sequential tunnelling in quantum dots or Pauli exclusion effects. However, in the tunnelling regime considered here ($t$~$10^{-6}$), the latter are minimal. Super-Poissonian noise on the other hand is a much rarer occurrence, as it always has to include some sort of interaction[11]. Experimentally, super-Poissonian noise was first observed in semiconductor double-wells and later in quantum dots[33]. Examples of super-Poissonian noise include bi-stable systems that lead to random telegraph noise in the transport, and coupling to inelastic modes[34]. Our data allows us to exclude all these mechanisms as the observed noise is bias dependent and asymmetric.

Instead, our data suggests the following scenario known from double quantum dots. Two tunnelling processes are present, one fast, accounting for almost all the tunnelling current, and one slow, acting as a switch for the first one. This switching mechanism is usually based on Coulomb interaction. For example, if the state of the slow process is occupied, it raises the energy level of the state necessary for the fast process and effectively block it, as shown in Fig 3. This leads to an effective switching i.e. bunching of electrons that causes super-Poissonian noise. Such mechanisms have been discussed in detail for mesoscopic systems[11,32,35] and confirmed by experiments in double quantum dots[36]. We note that this scenario *(i)* predicts a clear threshold energy after which the noise increases, *(ii)* is asymmetric with respect to energy, *(iii)* is localized on the atomic scale. The key insight that follows from our observations is that some slow process is involved, i.e. a form of charge trapping that is known from strongly polarizable insulators but not from a metal.

One expects these noise centers to correspond with some form of polarons being localized at defects in the crystal lattice. To shed light on their precise nature, we searched for impurity states, following earlier work[37,38] but now taking fully normalized $\frac{dI/dV}{I/V}$ density of states maps over a large energy range to be able to differentiate impurity states that overlap in energy. The energy dependent *dI/dV / (I/V)* maps reveal the spatial distribution of the different impurity states. Such states have been identified as (i) apical oxygen vacancies, (ii) Sr(Ca)-site impurities, (iii) interstitial oxygen dopants and (iv) Pb impurities.

Figure 4 shows signatures of these specific states; details can be found in the supplementary information.

Most importantly, we find a correspondence between positive-energy noise centers and the +1.1 V impurity state, as shown in Figure 4c-d. This impurity state has previously been identified as a apical oxygen vacancy[37] which, in the insulating oxygen materials, has various charge stabilized by lattice distortions. This amounts to strong evidence for this signature to be associated with the charge trapping process. The positive noise centers show all the signatures of an impurity state in the charge reservoir layer through which tunneling occurs but modulated by a slow charge trapping process. Surprisingly, we do not see noise centers at all impurity locations near impurity resonances at positive energy, nor do we observe impurity resonances at the site of the negative noise centers. This might be because the state is dark due to filtering mechanisms[30], or because of stronger coupling to the CuO layer.

In summary, we have presented direct evidence for the existence of slow charge trapping processes at defect sites in the form of the localized super-Poissonian noise signals. These are reminiscent of Coulombic impurities that occur generically in polarizable insulators, despite the three-dimensional superconducting state present in our samples. The c-axis physics of the cuprates appears to be in a quite literal sense similar as to what is found in e.g. the Al-$Al_2O_3$-Al barriers employed for Josephson junctions: No coherent charge transport is possible for normal electrons, while the virtual tunnelling of Cooper pairs suffices for a coherent Josephson contact. In the cuprates this is manifested of the form of the c-axis Josephson plasmons observed in the optical conductivity,[39] emerging from a completely overdamped charge transport in the normal state.[4] The surprise is that apparently an oxidic layer that is only two atoms thick living in a metallic environment suffices for the polaronic trapping of charge.

The role of the 'c-axis phenomena' in the mechanism of high-temperature superconductivity is a long standing question,[6,7,40,39] as is the unusual nature of the coupling of the polar insulator phonons to the electrons[6,7,9,10,40,41]. This acquired new impetus recently with the discovery that when a single layer of FeSe is removed from bulk and put on an polarizable insulator, the critical temperature increases by a factor of four[42], with evidence reported suggesting that the coupling to the polarizable insulating substrate may indeed play a critical role.[43] Further, interfaces of the polaronic insulators $SrTiO_3$ and $LaAlO_3$ host two dimensional superconductors with the highest $T_c$ per charge density ratios[41]. So much is clear from our findings that even for the atomically thin insulating layers the polar electron-phonon interactions are of a severity sufficing to slow down electronic motions to macroscopic time scales. Although a great challenge for established theory, this conundrum deserves further close consideration.

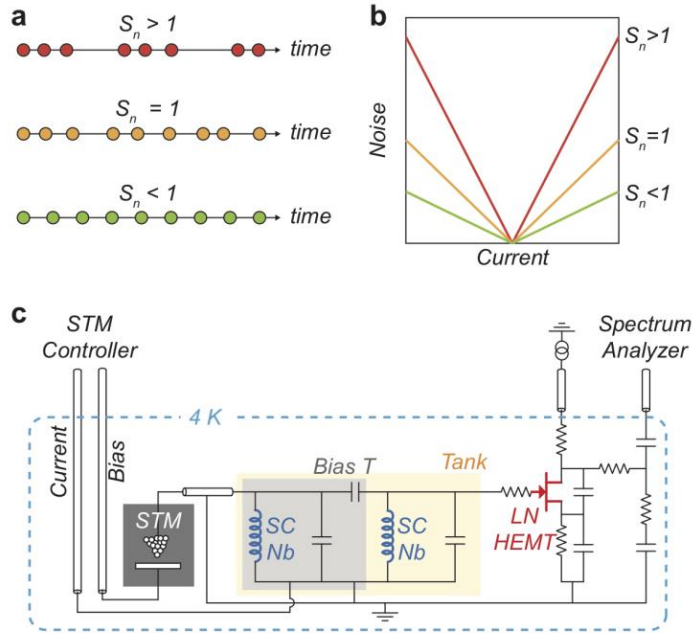

**Figure 1.** Scanning tunnelling noise spectroscopy as a new diagnostic tool. **a.** The classical flow of uncorrelated particles with charge q is a pure Poissonian process, S = 2q|I|. In the case of bunching, S > 2q|I| and we refer to the noise as super-Poissonian (red). In the case of anti-bunching, S < 2q|I| which is called sub-Poissonian (green). **b.** Noise S as a function of current in the cases described in (a). **c.** Measurement circuit that allows for spatial mapping of the current fluctuations in the STM junction. A bias-tee (indicated by light grey area) separates the low and high frequency components of the signal. Superconducting Niobium (Tc=9.2K, indicated in blue) inductors are used for the tank circuit (light orange box) that in combination with the custom build low-noise HEMT (LN HEMT in red) form the low-temperature amplification chain for noise measurements.

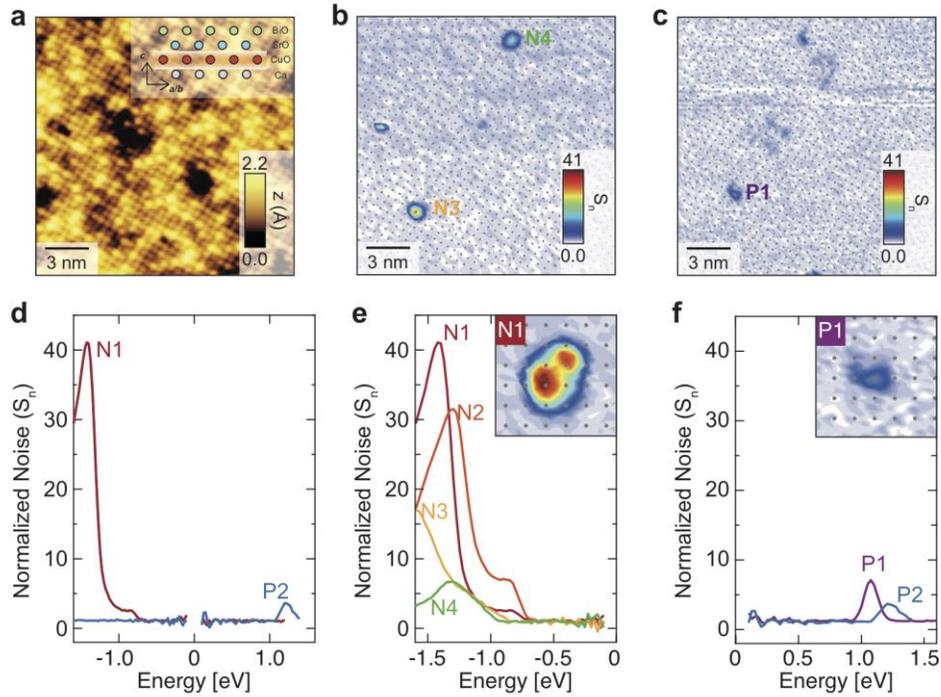

**Figure 2.** Observation of super-Poissonian noise centers. **a,** An atomic-resolution STM image of the BiO terminated overdoped bi-layer cuprate BiPbSrCaCuO surface in a 18.3 nm field of view (sample bias $V_B$ = -0.10 V, current $I_S$ = 0.1 nA. Pb substitution for Bi are visible as bright protrusions. **b, c,** Spatially resolved, background line-subtracted, noise maps at -1.2 (b) and +1.1 (c) eV in the same field of view as figure 2a. Most locations show homogeneous Poissonian noise ($S_n$=1), but a few atomic locations reveal striking enhancements. N3, N4, and P1 indicate the negative and positive noise centers. Grey dots represent the Cu lattice sites. **d,** Representative noise spectra on the atomic locations that exhibit super-Poissonian noise show the strong asymmetry. **e, f,** Noise spectra on various positive and negative noise centers. Each inset shows the spatial distribution of the noise enhancements (see also supplementary information).

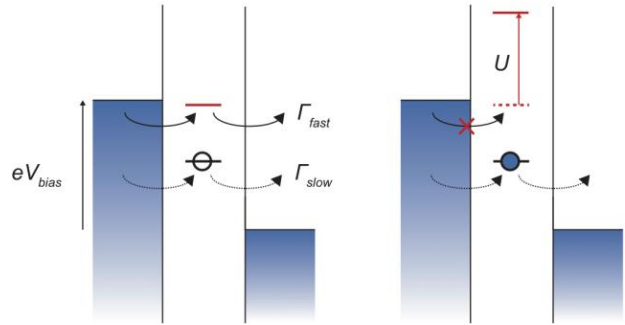

**Figure 3.** Example of modulated transport by slow charge trapping processes. Energy diagram of the co-tunnelling process via impurity states. Two tunnelling processes are possible; one of them has higher transmission rate than another and dominates the total tunnelling current. Since they are strongly coupled by Coulomb energy (*U*), the tunnelling through the higher impurity level is prohibited by charge trapping of the lower one

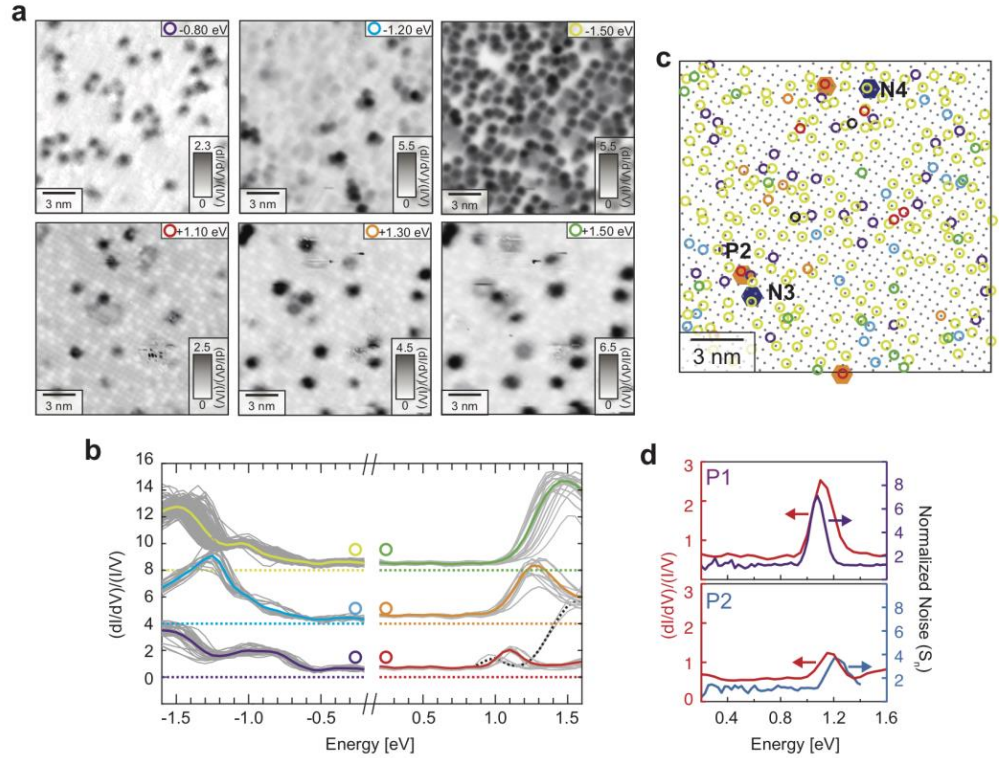

**Figure 4.** Bias-dependent conductance maps to identify impurity states and correlation with noise centers. **a**, Density of states at different energy levels: -1.5, -1.2, -0.8, +1.1, +1.3, and +1.5 eV. They were acquired in the same field of view in Fig. 2a. The enhancement in normalized differential conductance ((dI/dV)/(I/V)) reveals the spatial distribution of various impurity states. **b**, Normalized differential conductance spectra taken on the different impurity states in Fig. 3a. The thick colored lines represent the average of the individual gray spectra. **c,** Overview of all impurity states in this field of view. Grey dots display the Cu-lattice. Various symbols corresponding to the different impurity states classified based on (a-g). The noise centers are indicated by the blue (negative) and red (positive) hexagons. **d,** Correspondence between positive-energy noise center and the +1.1 V impurity state. The bottom panel represents the normalized noise spectrum on the noise center, the top panel shows the normalized dI/dV spectrum at the same location. P1 and P2 correspond to the positive noise centers shown in Fig. 2.